\documentclass[fleqn,twoside]{article}
\usepackage{espcrc2}

\usepackage{graphicx}
\usepackage[figuresright]{rotating}

\title{BEPCII and BESIII}

\author{Frederick A. Harris \address{Dept. of Physics and Astronomy, \\
        The University of Hawaii, \\ 
        Honolulu, HI 96822, USA} \\
For the BES Collaboration}

\begin{document}

\begin{abstract}
The Beijing Electron Collider is being upgraded (BEPCII) to a two-ring
collider with a design luminosity of $1 \times 10^{33}$cm$^{-2}$
s$^{-1}$ at a center-of-mass energy of 3.89 GeV. It will operate
between 2 and 4.2 GeV in the center of mass.  With this luminosity,
the BESIII detector will be able to collect, for example, 10 billion
$J/\psi$ events in one year of running.  This will be a unique
facility in the world opening many physics opportunities. BEPCII and
BESIII are currently under construction, and commissioning of both is
expected to begin in summer 2007.

\vspace{1pc}
\end{abstract}

\maketitle

\section{Introduction}

The Beijing Electron-Positron Collider (BEPC) at the Institute of High
Energy Physics (IHEP) in Beijing has been, until recently, a unique
facility running in the tau-charm center-of-mass energy region from 2
to 5 GeV with a luminosity at the $J/\psi$ peak of $5 \times
10^{30}$cm$^{-2}$ s$^{-1}$.  The Beijing Spectrometer
(BESI~\cite{besi} and BESII~\cite{besii}) detector at the BEPC has
operated since about 1990 and studied many physics topics, including a
precision measurement of the tau mass~\cite{taumass} and a detailed
R-scan~\cite{rscan}, and obtained 58 million events at the $J/\psi$,
14 million at the $\psi^{'}$, and over 30 pb$^{-1}$ at the $\psi{''}$.

 In 2003, the Chinese Government approved the upgrade of the BEPC to a
two-ring collider (BEPCII) with a design luminosity approximately 100
times higher than that of the BEPC.  This will allow unprecedented
physics opportunities in this energy region and contribute to
precision flavor physics.  In this paper, BEPCII and BESIII will be
described, along with their status.  Another paper presented at this
conference will provide more detail on BESIII physics and on BESIII
simulation studies~\cite{haibo}.

Currently BESII has been removed from the interaction region.  The
BEPCII linac installation is complete and that of the storage ring
started in March 2006. The installations of the muon counter and
the super-conducting magnet into the BESIII detector are also
complete.

\section{BEPCII}

BEPCII is a two-ring $e^+e^-$ collider that will run in the tau-charm
energy region ($E_{cm} = 2.0 - 4.2$ GeV) with a design luminosity of
$1\times 10^{33}$ cm$^{-2}$s$^{-1}$ at a beam energy of 1.89 GeV, an
improvement of a factor of 100 with respect to the BEPC. This is
accomplished by using multi-bunches and micro-beta. The upgrade
uses the existing tunnel and some old magnets. 

\begin{figure}  \centering
   \includegraphics*[width=0.45\textwidth]{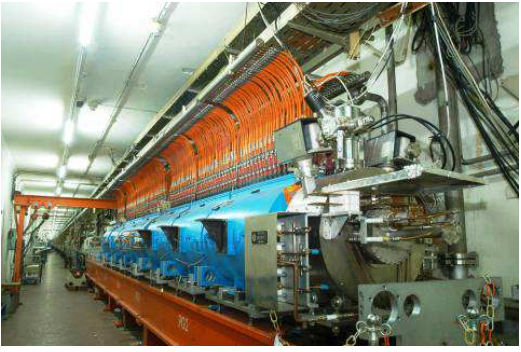}  
  \caption{\label{linac}The completed LINAC of BEPCII.
    }
 \end{figure}

The 2024 meter long linac has been upgraded with new klystrons, a new
electron gun, and a new positron source to increase its energy and
beam current; it can accelerate electrons and positrons up to 1.89 GeV
with an positron injection rate of 50 mA/min. Its installation
was completed in the summer of 2005 (see  Fig.~\ref{linac}), and it reached
most design specifications.

There are two storage rings with lengths of 237.5 meters.  The
collider has new super-conducting RF cavities, power supplies, and
control; super-conducting quadrupole magnets; beam pipes; magnets and
power supplies; kickers; beam instrumentation; vacuum systems; and
control.  The old dipoles are modified and used in the outer ring.
Electrons and positrons will collide at the interaction point with a
horizontal crossing angle of 11 mrad and bunch spacing of 8 ns.  Each
ring has 93 bunches with a beam current of 9170 mA.  The machine will
also provide a high flux of synchrotron radiation at a beam energy of
2.5 GeV.  The manufacture of major equipment such as magnets,
superconducting RF cavities (with the cooperation of KEK and MELCO)
and quadrupole magnets (with the cooperation of BNL), as well as the
cryogenics system, is complete. Installation in the tunnel began in
March 2006.  Beam collisions are expected in summer of 2007.

\section{BESIII}
\begin{figure}  \centering
   \includegraphics*[width=0.45\textwidth]{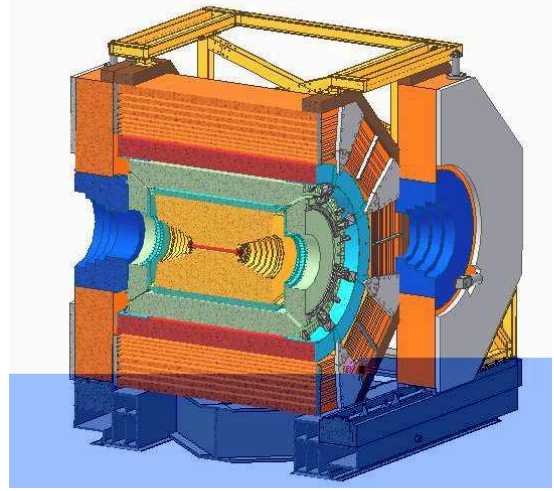}  
  \caption{\label{schematic}Schematic view of the BESIII detector.
    }
 \end{figure}

The BESIII detector consists of a berylium beam pipe, a helium-based
small-celled drift chamber, Time-Of-Flight counters for particle
identification, a CsI(Tl) crystal calorimeter, a super-conducting
solenoidal magnet with a field of 1 Tesla, and a muon identifier using
the magnet yoke interleaved with Resistive Plate Counters
(RPC). Fig.~\ref{schematic} shows the schematic view of the BESIII
detector, including both the barrel and endcap portions.
    
\subsection{Main Drift Chamber}    
\begin{figure}  \centering
   \includegraphics*[width=0.45\textwidth]{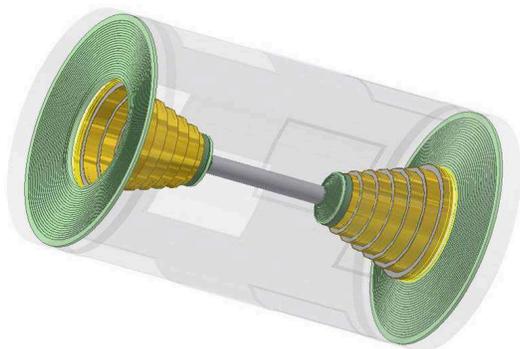}  
  \caption{\label{mdc}Schematic view of the MDC.
    }
 \end{figure}

The main drift chamber (MDC) is 2.58 meters in length and has an inner
radius of 63 mm and an outer radius of 0.81 m.  The inner and outer
cylinders are carbon fiber. As shown in Fig.~\ref{mdc}, there is a
short inner portion near the beam pipe, a stepped region, and a
cone shaped outer region.  The polar angle coverage is $\cos \theta =
0.83 $ for a track passing through all layers, and $\cos \theta = 0.93
$ for one that passes through 20 layers.  The endplates are machined
with a hole position accuracy better than 25 microns. Altogether there
are 43 layers of 25 micron gold plated tungsten wires; the field wires
are 110 micron gold-plated aluminum.  The cells are approximately
square, and the size of the half-cell is 6 mm in the inner portion of
the drift chamber and is 8.1 mm in the outer portion.  The chamber
will use a 60/40 He/$C_3H_8$ gas mixture.

The expected spatial, momentum, and $dE/dx$ resolutions are $\sigma_s
= 130 \mu$m, $\sigma_p/p= 0.5 \% $ at 1 GeV/$c$, and
$\sigma_{dE/dx}/dE/dx \sim 6 \%$, respectively.  Beam tests performed
with prototype electronics at KEK in a 1 T magnetic field yielded a
spatial resolution better than 110 microns and $dE/dx$ resolution
better than 5\%. The readout uses the CERN HPTDC. The wiring of the
drift chamber has been completed, and the electronics assembly of the
chamber has started.

\subsection{TOF}

Outside the MDC is the time of flight (TOF) system, which is crucial
for particle identification. It consists of a two layer barrel array
of 88 50 mm x 60 mm x 2320 mm BC408 scintillators in each layer and
endcap arrays of 48 fan shaped BC404 scintillators.  Hamamatsu R5942
fine mesh phototubes will be used - two on each barrel scintillator
and one on each endcap scintillator.  Expected time resolution for
kaons and pions and for two layers is 100 to 110 ps, giving a $2 \sigma$
$K/\pi$ separation up to 0.9 GeV/c for normal tracks.  This has
been confirmed in a beam test of a TOF counter using prototype
electronics.  The scintillator and phototubes for the TOF system will
be delivered before summer 2006, and the system will be tested and
ready for installation in January 2007. A laser TOF calibration system
is being built by the University of Hawaii.

\subsection{Calorimeter}
\begin{figure}  \centering
   \includegraphics*[width=0.30\textwidth]{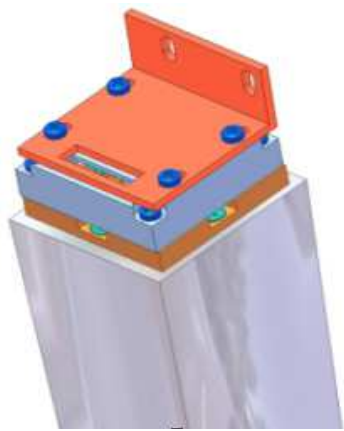}  
  \caption{\label{crystal}Schematic of the photodiode and preamp
    assembly on the end of a crystal.
    }
 \end{figure}

The CsI(Tl) crystal calorimeter contains 6240 crystals.  The typical
crystal is 5 $\times$ 5 cm$^2$ on the front face and 6.5 $\times$ 6.5
cm$^2$ on the rear face with a length of 28 cm or 15 radiation
lengths.  Figure~\ref{crystal} shows a schematic of the assembly
containing an aluminum plate with two photodiodes (Hamamatsu S2744-08)
with 10 mm by 20 mm sensitive area and an aluminum box for the preamp
mounted on the end of a crystal.  The expected energy and spatial
resolutions at 1 Gev are 2.5 \% and 0.6 cm, respectively.

The CsI(Tl) crystals are being produced at Saint-Gobain, Shanghai
Institute of Ceramics, and Hamamatsu (Beijing). Most of the crystals
have been delivered, and their size, light yield, uniformity, and
radiation hardness are satisfactory. A beam test shows that the
electronics noise from the preamplifier, main amplifier, charge
digitizer, and 18 meters of cable was less than 1000 electrons
equivalent per crystal, corresponding to about 220 keV of energy.

\begin{figure}  \centering
   \includegraphics*[width=0.45\textwidth]{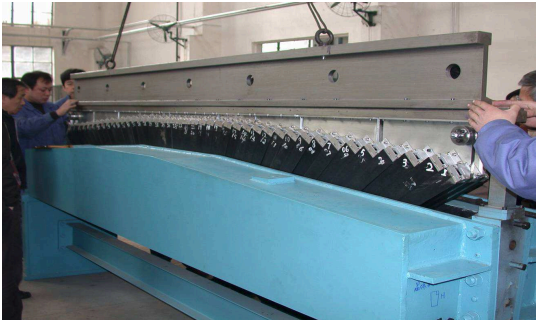}  
  \caption{\label{calorimeter} Prototype of the mechanical structure for
mounting the crystals.
    }
 \end{figure}

Figure~\ref{calorimeter} shows the prototype mechanical structure for
mounting the crystals.  The crystals are held by screws and there are
no walls between crystals.  Mechanical assembly will start soon and
should be completed by the end of the year.  By the end of the
year, all electronics boards should be tested and installed.

\subsection{Magnet}
\begin{figure}  \centering
   \includegraphics*[width=0.45\textwidth]{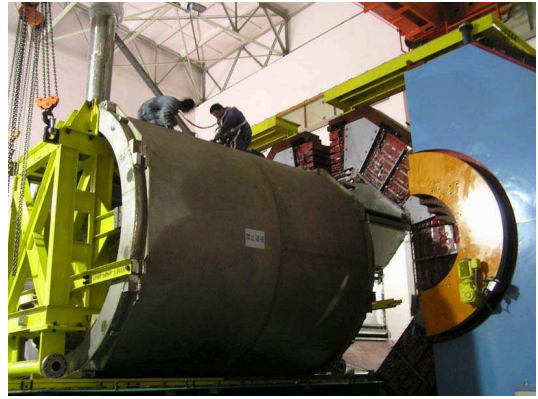}  
  \caption{\label{magnet}The super-conducting magnet during the
installation into the BESIII detector.  .  }
 \end{figure}

The BESIII super-conducting magnet is the first of its kind built in
China. The vacuum cylinder and the supporting cylinder are made in
China, in collaboration with the Wang NMR company of California, and
the wiring of the super-conducting cable and later the epoxy curing,
assembly, and testing were all done at IHEP with advice from experts
all over the world.  The superconducting magnet is a 3.91 m long
single layer solenoid with a 1 T magnetic field at a nominal current
of 3650 A. The field in the MDC will have a uniformity better than 5
\%, and it will be measured with an accuracy better than 0.25 \%. The
magnet is complete, and Fig.~\ref{magnet} shows its installation into
BESIII.  The magnet will be tested before summer, and the field
mapping will be done  with
the super conducting quadrupoles in place using a computer controlleed mapping machine before December 2006.

\subsection{Muon Counter}

The magnet return iron has 9 layers of Resistive Plate Chambers (RPC)
in the barrel and 8 layers in the endcap to form a muon counter.  The
electrodes of the RPCs are made from a special phenolic paper laminate
on bakelite, which has a very good surface quality. The gas used is Ar
: $C_2H_2F_4$ : Isobutane (50:42:8).  Extensive testing
and long term reliability testing has shown that the chambers have
high efficiency, low dark current, and good long term stability.

\begin{figure}  \centering
   \includegraphics*[width=0.45\textwidth]{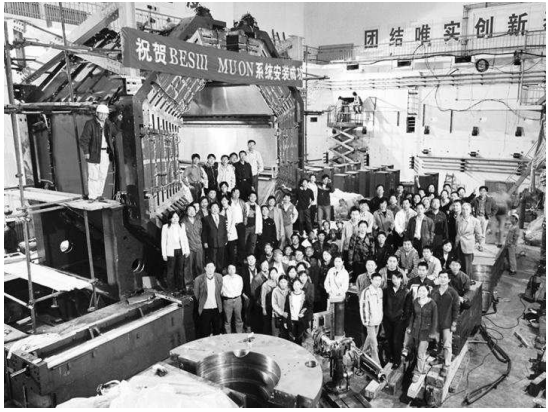}  
  \caption{\label{muon} BESIII group after successful installation of
RPCs for the muon counter. }
 \end{figure}

All the RPCs for the muon identifier have been manufactured, tested,
and installed, as shown in Fig.~\ref{muon}. The average dark current
and noise level for all chambers installed after one week's training
is 1.6 $\mu$A/m$^2$ and 0.095 Hz/cm$^2$, respectively, for a high
voltage corresponding to an average efficiency of 95 \%.  The spatial
resolution obtained was 16.6 mm.
    
\subsection{Trigger, Data Acquisition, and Offline Software}

The trigger design is almost finalized; it is pipelined and uses
FPGAs.  Information from the TOF, MDC,  and muon counter will be
used. The maximum trigger rate at the $J/\psi$ will be about 4000 Hz
with a good event rate of about 2000 Hz.  All boards
should be tested and installed  by the end of 2006. 

The whole data acquisition system has been tested to 8 kHz for an
event size of 12 Kb, which is a safety margin of a factor of two.  The
expected bandwidth after level one is 48 Mbytes/s. The data
acquisition system is 1000 times the performance of BESII.
    
\begin{figure}  \centering
   \includegraphics*[width=0.43\textwidth]{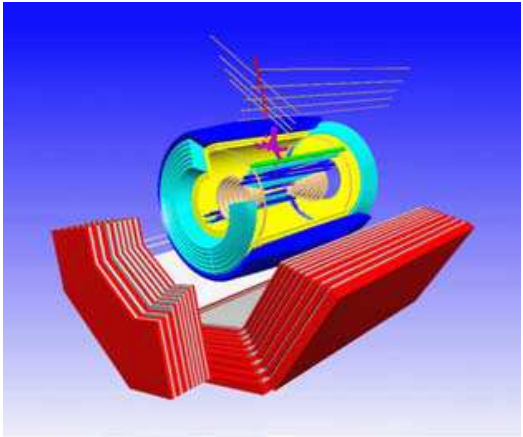}  
  \caption{\label{simulation1} BESIII event display.  }
 \end{figure}

\begin{figure}  \centering
   \includegraphics*[width=0.45\textwidth]{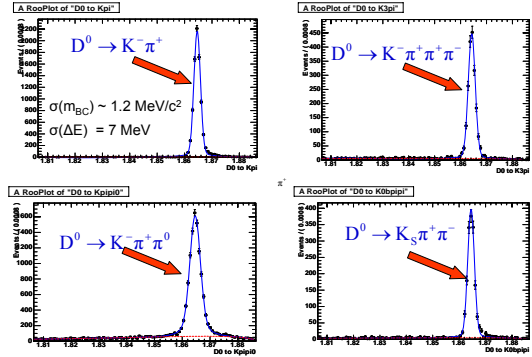}  
  \caption{\label{simulation2} Reconstructed $D^0$ events from a
  simulated  50,000 $\psi{''}$ inclusive event sample using BOSS.  }
 \end{figure}

The preliminary verson of the BES Offline Software System (BOSS) is complete.
A tremendous amount of work has been accomplished but much remains to
be done.  Simulation is based on Geant4.  Figure~\ref{simulation1}
shows a BESIII event display.
Figure~\ref{simulation2} shows $D$ decays
reconstructed from a simulated 50,000 $\psi{''}$ inclusive event
sample. The beam constrained mass resolution for $D^0 \to K^- \pi^+$
is $\sigma_{BC} = 1.2$ MeV/$c^2$.  Simulation and reconstruction is
done using BOSS.
    
\section{Physics in the tau-charm energy region}

The tau-charm energy region makes available a wide variety of interesting
physics.  Data can be taken at the $J/\psi$, $\psi(2S)$, and
$\psi(3770)$, at $\tau$ threshold, and at a energy to allow production
of $D_s$ pairs, as well as for an R-scan.  

In 2003, the construction of wiggler magnets at the CESR collider at
Cornell was approved that would allow the energy of their machine to
be lowered to compete with BES, and a test run was made with a single
wiggler magnet. Since 2005, CESR has had a full complement of these
magnets, and the luminosity of CLEOc is currently less than $1 \times
10^{32}$ cm$^{-2}$ s$^{-1}$.  CLEOc is scheduled to finish data taking
in 2008.

\begin{table*}[htb]
\caption{\label{3.1-1}
 Number of events expected for one year of running.}
\begin{center}
\begin{tabular}[!h]{| l | c | c | c | c| }
\hline
Physics & Center-of-mass  &  Peak          & Physics      & Number of \\ 
channel & Energy          & Luminosity     & cross    & Events per \\ 
        & (GeV)  & ($10^{33}$ cm$^{-2}$ s$^{-1}$) & section (nb) & year\\\hline
$J/\psi$   &   3.097  &  0.6  &  $\sim 3400$ & $10\times 10^9$ \\
$\tau$     &   3.67   &  1.0  &  $\sim2.4$   & $12 \times 10^6$ \\
$\psi(2S)$ &   3.686  &  1.0  &  $\sim640$   & $3.0 \times 10^9$ \\
$D$        &   3.770  &  1.0  &  $\sim5$     & $25 \times 10^6 $\\
$D_s$      &   4.030  &  0.6  &  $\sim0.32$  & $1.0 \times 10^6$ \\
$D_s$      &   4.140  &  0.6  &  $\sim0.67$  & $2.0 \times 10^6$ \\
\hline \end{tabular}
\end{center}
\end{table*}

BEPCII and BESIII will begin commissioning in summer 2007.
The design luminosity of BESIII is $1 \times 10^{33}$ cm$^{-2}$.
Clearly BESIII with higher luminosity will contribute greatly to precision
flavor physics: $V_{cd}$ and $V_{cs}$ will be measured with a
statistical accuracy of 1.6\%. $D^0 D^0$  mixing and CP violation will
be searched for. Table~\ref{3.1-1} gives the numbers of events
expected during one year of running at various energies.  Huge
$J/\psi$ and $\psi(2S)$ samples will be obtained.
The $\eta_C$, $\chi_{CJ}$, and $h_C$ can be studied with high
statistics.  The $\rho \pi$ puzzle will be studied with better
accuracy.   
Table~\ref{table1} shows a comparison of the BESIII
and CLEOc detectors.

\begin{table}[htb]
\begin{center}
\caption{Comparison of BESIII and CLEOc.}
\label{table1}
\begin{small}
\begin{tabular}{|l|c|c|}
\hline
Detector        & BESIII & CLEOc \\ 
\hline
MDC             & $\sigma_{xy} = 130 \mu$m           & 90 $\mu$m \\
                & $\Delta P/P = 0.5 \%$ (@1 GeV/c)     & 0.5 \%    \\
                & $\sigma_{dE/dx} = 6\%$             & 6 \%       \\ \hline
EMC             & $\Delta E/E = 2.5 \%$ (@1GeV) & 2.0 \% \\
                & $\sigma_z = 0.6/\sqrt{E}$ cm       &  $0.5/\sqrt{E}$
cm \\ \hline
TOF             & $\sigma_T = $100 - 110 ps    & Rich  \\
                &  Double layer                      &       \\ \hline 
$\mu$           &  9 layers                          & -- \\ 
counter         &                                    &  \\ \hline
Magnet          &  1 T                               & 1 T \\ \hline
\end{tabular}\\[2pt]
\end{small}
\end{center}
\end{table}

\section{BESIII Collaboration}

A BESIII collaboration meeting was held January 10 - 12, 2006 at IHEP in
Beijing, China.  More than one hundred collaborators from 21
institutions and six countries attended the meeting, including China,
United States, Japan, Sweden, Germany, and Russia.
This meeting was historic as the governance
rules of the collaboration were approved and used for the first
time. Under these rules, the Institutional Board (IB) was established
and its chair was elected.

A conference  CHARM2006 will be held in Beijing in June of this
year to discuss the physics potential of BESIII, and a US-China
workshop on BESIII collaboration will be held immediately afterwards.
All interested parties are welcomed.

\end{document}